# Image processing-based domain-matching simulation for heteroepitaxy

T. Ishiyama,[a)] T. Suemasu, and K. Toko[b)]

Institute of Applied Physics, Univ. of Tsukuba, 1-1-1 Tennodai, Tsukuba, Ibaraki 305-8573, Japan
[a)] ishiyama.takamits.ta@alumni.tsukuba.ac.jp, [b)] toko@bk.tsukuba.ac.jp

**Abstract**

Heteroepitaxy of functional thin films on single-crystal substrates is one of the most general themes in electronic materials research. Here, we propose an algorithm based on image processing for the rapid simulation of heteroepitaxial relationships. The superposition and rotation of various lattice plane images of the film and substrate, which were automatically generated from the crystal structure, rapidly verified all domain matching patterns. Furthermore, the comprehensive validation allowed us to discuss domain matching from multiple perspectives, such as mismatch, matching period, and density of matching lattice points. Therefore, the proposed algorithm will contribute to materials informatics, streamlining a wide range of materials research for functional thin films.

## 1. Introduction

Many functional thin films that sustain current electronics have been obtained using heteroepitaxy on single-crystal substrates. Because lattice matching is the key to the epitaxial growth of high-quality functional thin films, selecting substrates and their crystal orientations is important.[1] The selection is relatively simple when materials possess similar crystal structures and lattice constants, such as the relationship between germanium and III-V compound semiconductors.[2] However, such cases are rather uncommon. In fact, the heteroepitaxy of many functional thin films, such as silicides,[3,4] germanides,[5,6] nitrides,[7,8] has been achieved not by lattice matching but by "domain matching (DM)," a relationship that is an integer multiple of the unit lattice of difference crystal structures.

Determining the DM between materials with different crystal structures requires the examination of many patterns with different crystal orientations and integer multiples of the lattice constants and rotations. The manual simulation of DM for a single material requires significant effort. Therefore, in many cases, the DM is inductively derived from the experimental results. This approach is labor-intensive and also leads to missing the best material selection. Therefore, various methods have been proposed to predict DM from crystal structures (such as Zur and McGill's lattice-matching algorithm).[9–13] Moreover, programs have also been developed to perform DM simulations based on the mathematical formulas.[14–18] In recent years, the above algorithm was combined with density functional theory to discuss DM from energy stability.[19–25] These methods helped in simplifying and accelerating material selection in heteroepitaxy, and also discusses experimental results. Conversely, the discussion of the positional relationship between the lattice planes remains important in the interpretation of the DM suggested by these advanced simulations and in examining the possibility of epitaxial growth. Therefore, this study constructed a visual DM verification program based on image processing by the simple rotation and superposition of the lattice planes. The comprehensive verification of DM patterns allows us to derive multifaceted information such as mismatch, matching period, and density of matching lattice points.

## 2. Code description

Figure 1 shows the flowchart of the DM simulation algorithm proposed in this study. This method calculates the overlapping of lattice points for various combinations of crystal orientations and rotation angles, $\theta$. First, a lattice image was created by arranging lattice points. This arrangement is automatically generated by using the crystal structure, including lattice constants, as input. Then, the two lattice planes are superposed such that their origins coincide, and only the lattice plane of the film is rotated. Considering the symmetry of the two-dimensional Bravais lattice, a search range of 0°–90° for $\theta$ is sufficient. The sum of the direct products of the lattice images, namely, the area of the overlapping region $A$, is the leading indicator of DM, which is extremely simple, versatile, and intuitive. In addition, the algorithm can derive the area of the DM unit cell $C$ and the mismatch between the cell sizes. This information suggests candidates for the best DM from multiple perspectives.

We used the method shown in Algorithm 1 to perform an image-processing-based DM search. The generated_lattice_image is a function that generates a lattice image (Image), considering the lattice constants $L$, Miller indexes $m_S$ and $m_F$, size of the lattice points (point_size) $k$, width $W$, and height $H$ of the generated image as variables. The subscripts S and F represent the substrate and film, respectively. The pixel value for drawing lattice points was set to 1, and 0 otherwise. Using this setup, $A$ can be quickly obtained by simply computing the direct product of the images and summing them. The DM verification assumes that

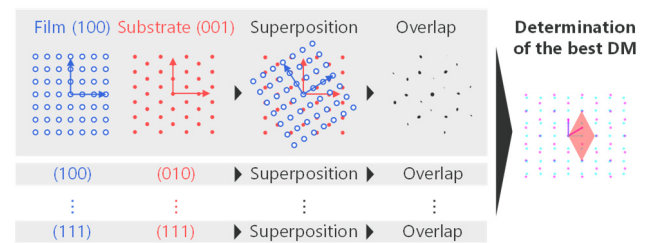

**Figure 1.** Schematic of the proposed algorithm.

**Notation List:**
- substrate: S
- film: F
- lattice constants: L
- a set of Miller indices: m
- Miller index search space: M
- angle of rotation: $\theta$
- overlapping area: A
- pixel coordinates: x, y
- lattice point size: k
- image width: W
- image height: H

```
Input: S, F
A_max = 0
θ_max = 0°
For m_S in M_S do
    For m_F in M_F do
        Image_S = generate_lattice_image(L_S, m_S, k)
        Image_F = generate_lattice_image(L_F, m_F, k)
        For 0° ≤ θ ≤ 90° do
            Image_Fθ = rotate_image(Image_F, θ)
            A = Σ_(x,y) Image_S(x,y)·Image_Fθ(x,y)
            If A > A_max do
                θ_max = θ
                A_max = A
                matching = (m_S, m_F)
        End For
    End For
End For
Return matching, θ_max, A_max
```

**Algorithm 1.** Pseudocode for a program to determine the combination of crystal orientations and rotation angles $\theta$ that yields the DM.

one point of the smallest parallelograms in the DM region (DM cell) is correctly placed because the degree of DM (e.g., the size of the DM cell and lattice misalignment) can be determined by the placement of the remaining three points. Therefore, one lattice point was always set at the center of the image. The rotate_image function rotates the input image by an angle $\theta$ from the image center. Image$_{F\theta}$ is the film lattice image Image$_F$ rotated by an angle $\theta$ by rotate_image. For each combination of lattice planes, $A$ is a function of the angle $\theta$ and conditions under which the lattice image is generated, namely, $k$, $W$, and $H$. The overlapping regions of lattice points in this operation can be created if $k$ is sufficiently large. When $W$ and $H$ are large, the lattice consistency can be verified for long periods, and when $k$ is small, the tolerance for considering coincident lattice points can be tightened. Thus, DM can be verified from multiple angles according to the objective by adjusting the parameters. This algorithm was used to determine the combination of $m$ and $\theta$, $\theta_{max}$ when $A$ attained its maximum value. This search is simple and involves only three For loops. Because there is no complex calculation process in between, the DM of the substrate and film can be rapidly simulated.

## 3. Applications

Using Ge and FeGe as examples, we examined the DM and verified the effectiveness of the proposed method. Figure 2(a) shows $A(\theta)$ for various combinations of the lattice planes. For the Ge(100)//FeGe(111) and Ge(110)//FeGe(100) combinations, symmetries can be observed at $\theta$ and 90° − $\theta$, which are because of the inversion and four-fold symmetry of the two-dimensional lattice. Among these, the Ge(110)/FeGe(110) combination with $\theta$ = 35.2° yielded the largest $A(\theta)$ = 2511 px. Figures 2(b) and (c) show the superpositions of the Ge(110)//FeGe(110) lattice images and DM cells at $\theta$ = 35.2° and 74.3°, respectively, where $A(\theta)$ peaks. It can be seen that $\theta$ = 35.2° produces more consistent DM than $\theta$ = 74.3°, considering that the smaller area of the DM cell indicates that the lattice points are coincident at a high density. Here, the angular step was $\Delta\theta$ = 0.1°, and there were nine combinations of crystal planes. This setup corresponds to 8100 DM verifications. Despite the large number of trials, the total processing time was 86 s. The calculations were completed in such a short period of time because the algorithm comprised simple operations such as NumPy array generation, cv2 rotation, and direct product.

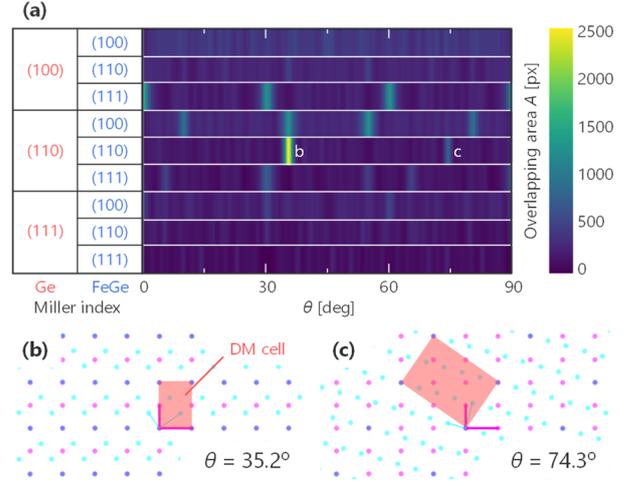

**Figure 2.** DM simulation for Ge and FeGe. (a) Heatmap summarizing $A$ as a function of $\theta$ for various combinations of lower-order planes. Ge(110)//FeGe(110) alignment relation for lattice images with $\theta$ = (b) 35.2° and (c) 74.3°. The dots indicate the lattice points of Ge (pink), FeGe (light blue), and overlapping (dark blue). Here, $H = W = 1000$ px and $k = 3$ px.

Table 1 lists the details of the DM of Ge and FeGe in the $A(\theta)$ peak shown in Figure 2. We note that the result is an example because it depends on the parameters $H$, $W$, and $k$. Let $a$ and $b$ be the lengths of the two sides of the DM unit cell and $\Delta a$ and $\Delta b$ be the mismatches on each side. In the case with the largest $A$, Ge(110)//FeGe(110) at $\theta$ = 35.2°, $C$ exhibited the smallest value of 91 Å². However, for Ge(110)//FeGe(100) at $\theta$ = 9.8° and 80.2°, which yielded the smallest mismatch ratio, $C$ was relatively large. One of the advantages of the proposed algorithm is that several DM patterns can be derived immediately, which helps in considering various possibilities.

**Table 1.** Simulation results for the DM of lower-order planes in Ge and FeGe for $H = W = 1000$ px and $k = 3$ px.

| Plane | | DM | | | Ge | | FeGe | | Mismatch | |
|---|---|---|---|---|---|---|---|---|---|---|
| Ge | FeGe | θ [deg] | A [px] | C [Å²] | a [Å] | b [Å] | a [Å] | b [Å] | Δa [%] | Δb [%] |
| (100) | (111) | 0 | 1409 | 449 | 11.32 | 39.61 | 11.39 | 39.44 | 0.62 | -0.43 |
| | | 30 | 1295 | 449 | 11.32 | 39.61 | 11.39 | 39.44 | 0.62 | -0.43 |
| | | 60 | 1295 | 449 | 11.32 | 39.61 | 11.39 | 39.44 | 0.62 | -0.43 |
| | | 90 | 1409 | 449 | 11.32 | 39.61 | 11.39 | 39.44 | 0.62 | -0.43 |
| (110) | (100) | 9.8 | 950 | 458 | 19.60 | 23.33 | 19.72 | 23.24 | **0.61** | **-0.39** |
| | | 35.2 | 1313 | 133 | 9.80 | 13.86 | 9.30 | 13.95 | -5.10 | 0.65 |
| | | 54.8 | 1313 | 133 | 9.80 | 13.86 | 9.30 | 13.95 | -5.10 | 0.65 |
| | | 80.2 | 950 | 458 | 19.60 | 23.33 | 19.72 | 23.24 | **0.61** | **-0.39** |
| (110) | (110) | 35.2 | **2511** | **91** | 8.00 | 11.32 | 8.05 | 11.39 | 0.63 | 0.62 |
| | | 74.3 | 83.7 | 335 | 13.86 | 24.00 | 13.95 | 24.15 | 0.65 | 0.62 |

## 4. Conclusion

We constructed an algorithm based on image processing to simulate the heteroepitaxial relationship between the film and substrate. The superposition and rotation of the various lattice plane images of the film and substrate, which were generated automatically from the crystal structure, verified the entire DM pattern. For example, when analyzing the low-order crystal planes in Ge and FeGe, 8100 cases for DM were verified in only 86 s. Furthermore, its comprehensive verification allowed a multifaceted evaluation of DM: Ge(110)//FeGe(110) yielded the largest lattice point matching and smallest matching period, whereas Ge(110)//FeGe(100) exhibited the smallest mismatch. The proposed algorithm is expected to aid substrate material exploration and interpretation of experimental results in the heteroepitaxial growth of functional thin films, and therefore, significantly streamline the research and development of electronic materials.

## 5. Code availability

The Python program is available at https://github.com/narishiro/lattice_matching. It comprises three cells: (i) *Import & Functions*, (ii) *Lattice* and (iii) *Matching*. The *Import & Functions* define the functions necessary for Algorithm 1. The parameters $k$, $W$, and $H$ are also defined in this cell. The lattice constants of the substrate and film in the dictionary *crystal* are entered in the *Lattice* cell. The *simulate_matching* function in *Matching* can be executed to automatically try various crystal orientations and rotation angles to obtain the best DM pattern. The calculation results are stored in variable *out*. In addition, *cv2.imshow* can be used to illustrate the matching relationship between lattices.